\documentclass[12pt]{article}
\usepackage{epsfig}
\usepackage{a4p}

\begin{document}
\def\be{\begin{equation} }
\def\ee{\end{equation} }
\def\ba{\begin{eqnarray} }
\def\ea{\end{eqnarray} }
\def\ban{\begin{eqnarray*} }
\def\ean{\end{eqnarray*} }
\def\epem{\mbox{e}^+\mbox{e}^-}
\def\eegg{\epem\to\gamma\gamma}
\def\eeggg{\epem\to\gamma\gamma(\gamma)}
\def\ggg{\gamma\gamma(\gamma)}
\def\ct{\cos{\theta}}
\def\cte{\cos{\theta^{\ast}}}
\def\pl{p_{\rm l}}
\def\pt{p_{\rm t}}
\def\g{\gamma}
\def\B{{\mathcal{B}}}
\def\O{{\mathcal{O}}}
\def\R{{\mathcal{R}}}
\def\E{{\mathcal{E}}}
\def\Lpm{\Lambda_{\pm}}
\def\xmc{\left(\frac{d\sigma}{d\Omega}\right)_{\rm MC}}
\def\xb{\left(\frac{d\sigma}{d\Omega}\right)_{\rm Born}}
\def\xl{\left(\frac{d\sigma}{d\Omega}\right)_{\Lambda_{\pm}}}
\def\xq{\left(\frac{d\sigma}{d\Omega}\right)_{\rm \Lambda '}}
\def\xe{\left(\frac{d\sigma}{d\Omega}\right)_{\rm e^{\ast}}}
\def\xsn{\frac{d\sigma}{d\Omega}}
\def\ca{$I$}
\def\cb{$I\!I$}
\def\cc{$I\!I\!I$}
\def\dd{$I\!V$}
\def\acol{\xi_{\rm acol}}
\def\acop{\xi_{\rm aplan}}

\begin{titlepage}
\begin{center}{\Large   EUROPEAN LABORATORY FOR PARTICLE PHYSICS
}\end{center}\bigskip
\begin{flushright}
       CERN-EP/98-092  \\ 
       8 June 1998
\end{flushright}
\bigskip\bigskip\bigskip\bigskip\bigskip
\begin{center}{\LARGE\bf  
\boldmath
Multi-photon production in $\epem$ collisions \\[1ex]
at $\sqrt{s} =$ 183 GeV 
\unboldmath
}\end{center}\bigskip\bigskip
\begin{center}{\LARGE The OPAL Collaboration

}\end{center}\bigskip\bigskip

\begin{abstract}
The process $\eegg(\g)$ is studied using data recorded with the OPAL detector 
at LEP. The data sample corresponds to a total integrated luminosity of 
56.2 pb$^{-1}$ taken at a centre-of-mass energy of 183 GeV. 
The measured cross-section agrees well with the expectation from QED.  
A fit to the angular distribution is used to obtain improved limits at 
95\%\ CL on the QED cut-off parameters: $\Lambda_+ > $ 233 GeV and 
$\Lambda_- > $ 265 GeV
as well as a mass limit for an excited electron, $M_{\rm e^{\ast}} >$ 227 GeV
assuming equal $\rm e^{\ast}e\gamma$ and  $\rm ee\gamma$ couplings.
No evidence for resonance production is found in the invariant mass 
spectrum of photon pairs. Limits are obtained for the cross-section times 
branching ratio for a resonance decaying into two photons.
\end{abstract}
\vspace*{1cm}

\begin{center}

{\large
(To be submitted to Phys. Lett.)
}\end{center}
\end{titlepage}
\begin{center}{\Large        The OPAL Collaboration
}\end{center}\bigskip
\begin{center}{
K.\thinspace Ackerstaff$^{  8}$,
G.\thinspace Alexander$^{ 23}$,
J.\thinspace Allison$^{ 16}$,
N.\thinspace Altekamp$^{  5}$,
K.J.\thinspace Anderson$^{  9}$,
S.\thinspace Anderson$^{ 12}$,
S.\thinspace Arcelli$^{  2}$,
S.\thinspace Asai$^{ 24}$,
S.F.\thinspace Ashby$^{  1}$,
D.\thinspace Axen$^{ 29}$
% ...
}\end{center}\bigskip
\bigskip
$^{  1}$School of Physics and Astronomy, University of Birmingham,
Birmingham B15 2TT, UK
\newline
$^{  2}$Dipartimento di Fisica dell' Universit\`a di Bologna and INFN,
I-40126 Bologna, Italy
\newline
$^{  3}$Physikalisches Institut, Universit\"at Bonn,
D-53115 Bonn, Germany
\newline
$^{  4}$Department of Physics, University of California,
Riverside CA 92521, USA
\newline
$^{  5}$Cavendish Laboratory, Cambridge CB3 0HE, UK
\newline
$^{  6}$Ottawa-Carleton Institute for Physics,
Department of Physics, Carleton University,
Ottawa, Ontario K1S 5B6, Canada
\newline
$^{  7}$Centre for Research in Particle Physics,
Carleton University, Ottawa, Ontario K1S 5B6, Canada
\newline
$^{  8}$CERN, European Organisation for Particle Physics,
CH-1211 Geneva 23, Switzerland
\newline
$^{  9}$Enrico Fermi Institute and Department of Physics,
University of Chicago, Chicago IL 60637, USA
\newline
$^{ 10}$Fakult\"at f\"ur Physik, Albert Ludwigs Universit\"at,
D-79104 Freiburg, Germany
\newline
$^{ 11}$Physikalisches Institut, Universit\"at
Heidelberg, D-69120 Heidelberg, Germany
\newline
$^{ 12}$Indiana University, Department of Physics,
Swain Hall West 117, Bloomington IN 47405, USA
\newline
$^{ 13}$Queen Mary and Westfield College, University of London,
London E1 4NS, UK
\newline
$^{ 14}$Technische Hochschule Aachen, III Physikalisches Institut,
Sommerfeldstrasse 26-28, D-52056 Aachen, Germany
\newline
$^{ 15}$University College London, London WC1E 6BT, UK
\newline
$^{ 16}$Department of Physics, Schuster Laboratory, The University,
Manchester M13 9PL, UK
\newline
$^{ 17}$Department of Physics, University of Maryland,
College Park, MD 20742, USA
\newline
$^{ 18}$Laboratoire de Physique Nucl\'eaire, Universit\'e de Montr\'eal,
Montr\'eal, Quebec H3C 3J7, Canada
\newline
$^{ 19}$University of Oregon, Department of Physics, Eugene
OR 97403, USA
\newline
$^{ 20}$Rutherford Appleton Laboratory, Chilton,
Didcot, Oxfordshire OX11 0QX, UK
\newline
$^{ 22}$Department of Physics, Technion-Israel Institute of
Technology, Haifa 32000, Israel
\newline
$^{ 23}$Department of Physics and Astronomy, Tel Aviv University,
Tel Aviv 69978, Israel
\newline
$^{ 24}$International Centre for Elementary Particle Physics and
Department of Physics, University of Tokyo, Tokyo 113, and
Kobe University, Kobe 657, Japan
\newline
$^{ 25}$Institute of Physical and Environmental Sciences,
Brunel University, Uxbridge, Middlesex UB8 3PH, UK
\newline
$^{ 26}$Particle Physics Department, Weizmann Institute of Science,
Rehovot 76100, Israel
\newline
$^{ 27}$Universit\"at Hamburg/DESY, II Institut f\"ur Experimental
Physik, Notkestrasse 85, D-22607 Hamburg, Germany
\newline
$^{ 28}$University of Victoria, Department of Physics, P O Box 3055,
Victoria BC V8W 3P6, Canada
\newline
$^{ 29}$University of British Columbia, Department of Physics,
Vancouver BC V6T 1Z1, Canada
\newline
$^{ 30}$University of Alberta,  Department of Physics,
Edmonton AB T6G 2J1, Canada
\newline
$^{ 31}$Duke University, Dept of Physics,
Durham, NC 27708-0305, USA
\newline
$^{ 32}$Research Institute for Particle and Nuclear Physics,
H-1525 Budapest, P O  Box 49, Hungary
\newline
$^{ 33}$Institute of Nuclear Research,
H-4001 Debrecen, P O  Box 51, Hungary
\newline
$^{ 34}$Ludwigs-Maximilians-Universit\"at M\"unchen,
Sektion Physik, Am Coulombwall 1, D-85748 Garching, Germany
\newline
\bigskip\newline
$^{  a}$ and at TRIUMF, Vancouver, Canada V6T 2A3
\newline
$^{  b}$ and Royal Society University Research Fellow
\newline
$^{  c}$ and Institute of Nuclear Research, Debrecen, Hungary
\newline
$^{  d}$ and Department of Experimental Physics, Lajos Kossuth
University, Debrecen, Hungary
\newline
$^{  e}$ on leave of absence from the University of Freiburg

\section{Introduction}

This paper reports a study of the process $\eeggg$ using data 
recorded with the OPAL detector at LEP at an average centre-of-mass energy 
of 182.7 GeV 
with an integrated luminosity of 56.2 pb$^{-1}$. At LEP energies,
this is one of the few processes having negligible contributions from the weak 
interaction. Since the QED differential cross-section is precisely predicted by
theory \cite{drell,mcgen}, searches for deviations from the expected angular 
distribution are a sensitive probe
for non-standard physics processes contributing to these photonic final states.
Any non-QED effects described by the general framework of effective Lagrangian 
theory are expected to increase with centre-of-mass energy \cite{eboli}. 
A comparison of the measured photon angular distribution with the QED 
expectation can be used to place limits on the QED cut-off parameters 
$\Lambda_{\pm}$, contact interactions ($\epem\g\g$) and non-standard 
$\epem\g$-couplings as described in section 3. 
The possible existence of an excited electron, $\rm{e}^*$, which would also
change the angular distribution~\cite{litke}, is investigated. In addition, a 
search is made for the possible production of a resonance X via 
\mbox{$\epem\to$ X$\gamma$}, followed by the decay \mbox{X $\to \g\g$}, 
using the invariant mass spectrum of photon pairs in three-photon final states. 

The process $\eeggg$ has been analysed previously at lower energies 
\cite{ich,l3,aleph,delphi}. The main differences from the previous OPAL 
analysis at $\sqrt{s}$ = 130 -- 172 GeV \cite{ich} are an increased 
angular acceptance and an improved rejection of non-physics backgrounds.
The following section contains a brief description of the OPAL detector
and of the Monte Carlo simulated event samples.
Section 3 describes the QED differential cross-sections for $\eeggg$, as well
as those from several models describing extensions to QED.
In sections 4 -- 6 the analysis is described in detail. The results are
presented in section 7. 

\section{The OPAL detector and Monte Carlo samples}

A detailed description of the OPAL detector can be found in Ref. \cite{det}.
The polar angle,~$\theta$, is measured with respect to the electron-beam
direction and $\phi$ is the azimuthal angle.
For this analysis the most important detector component is the 
electromagnetic calorimeter (ECAL) which is divided into two parts, the barrel 
and the endcaps. The barrel covers polar angles with $|\ct|<0.82$ and 
consists of 9440 lead-glass blocks in a quasi-pointing geometry. The endcaps 
cover the polar 
angle range $0.81<|\ct |<0.98$ and each consists of 1132 blocks.
The spatial resolution is about 11~mm, corresponding to 0.2$^{\circ}$ in 
$\theta$. The energy resolution for high energy photons is about 1.6\%\ 
in the barrel and \mbox{3 -- 5\% } in the endcaps depending on the angle.
The ECAL surrounds the tracking chambers. Hit information from the 
jet chamber and the vertex drift chamber is used to reject events which are 
consistent with having charged particles coming from the interaction point. 
Outside the ECAL are the hadronic calorimeter (HCAL), which is incorporated 
into the magnet yoke, and beyond that the muon chambers. Both the HCAL and the 
muon chambers are used to reject cosmic ray events.

Various Monte Carlo samples are used to study the selection efficiency and
expected background contributions.
For the signal process $\eeggg$ the RADCOR \cite{mcgen} generator is used,
while FGAM \cite{fgam} is used for $\eegg\gamma\gamma$. FGAM does not take
into account the electron mass and hence can not be used if at least one 
photon is along the beam-axis.
The Bhabha process is simulated using BHWIDE \cite{mcbh} and TEEGG \cite{mcte}.
The processes $\epem\to\overline{\nu}\nu\g(\g)$,
$\mu^+\mu^-$ and $\tau^+\tau^-$ are simulated using
KORALZ \cite{mctt}. PYTHIA \cite{mcmh} is used for hadronic events. 
All samples were processed through the OPAL detector simulation program 
\cite{mcdet} and reconstructed in the same way as real data.

\section{Cross-section for the process \boldmath $\eegg$ \unboldmath}

The Born-level differential cross-section for the process $\eegg$,
in the relativistic limit of lowest order QED is given by \cite{bg}:
\be
\xb = \frac{\alpha^2}{s}\;\frac{1+\cos^2{\theta} }{1-\cos^2{\theta} } \; ,
\label{born}
\ee
where $s$ denotes the square of the centre-of-mass energy, $\alpha$ is the
fine-structure constant and the event angle $\theta$ is the polar angle of one 
photon. Since the two photons cannot be distinguished, the event angle is 
defined such that $\ct$ is positive. 

In Ref. \cite{drell} possible deviations from the QED cross-section for Bhabha 
and M{\o}ller scattering are pa\-ra\-me\-tri\-zed in terms of cut-off parameters
$\Lambda_\pm$. 
These parameters correspond to a short-range exponential term added 
to the Coulomb potential. This ansatz leads to a modification of the 
photon angular distribution of the form 
\be
\xl   =  \xb \; \left[
1\pm\frac{s^2}{2\Lambda_\pm^4}\sin^2{\theta} \right] .
\label{lambda} 
\ee
Alternatively, in terms of effective Lagrangian theory, a gauge-invariant operator may be added to 
QED. Depending on the dimension of the operator, different deviations from QED 
can be formulated \cite{eboli}. 
Contact interactions ($\g\g\epem$) or non-standard $\g\epem$ couplings 
described by dimension 6, 7 or 8 operators lead to angular distributions 
with different mass scales $\Lambda$. In most cases these deviations are
functionally similar \cite{ich}. The cross-section predicted by a 
\mbox{dimension-7} Lagrangian is given by
\ba
\xq & = & \xb + \frac{s^2}{32\pi}\frac{1}{\Lambda'^6} \; .
\label{qed7} 
\ea
 
The existence of an excited electron ${\rm e}^*$ with an
${\rm e}^*{\rm e}\g$ coupling
would contribute to the photon production process via $t$-channel exchange.
The resulting deviation from $\xb$ depends on the ${\rm e^*}$ mass $M_{\rm e^*}$ and
the coupling constant $\kappa$ of the $\mathrm{e^* e\g}$ vertex relative 
to the $\mathrm{ee\g}$ vertex \cite{litke}:
\be \label{estar}
\left(\frac{d\sigma}{d\Omega}\right)_{\rm e^*} = 
\xb + f(M_{\rm e^{\ast}},\kappa ,s,\ct ).
\ee
The function $f$ is explicitly given in Ref. \cite{ich}. 
In the limit $ M_{\rm e^*} \gg \sqrt{s}$, $\xe$ approaches $\xl$ with the
mass related to the cut-off para\-meter by 
$ M_{\rm e^*} = \sqrt{\kappa}\;\Lambda_+$.

In a multi-photon event, it is important to choose an appropriate definition of
the event angle. The event angle is not uniquely defined since the two
highest energy photons in general are not exactly back-to-back.
The event angle $\cte$ used in this paper is defined as
\be 
\cos{\theta^{\ast}} = \left|\sin{\frac{\theta_1 - \theta_2}{2}}\right|
     \; {\Bigg /} \; \left( {\sin{\frac{\theta_1 + \theta_2}{2}}}\right) \; ,
\label{ctstar} \ee
where $\theta_1$ and $\theta_2$ are the polar angles of the two highest energy
photons. This definition was chosen such that the deviations in the angular 
distribution with respect to the Born level are small, in this case between 
3 -- 6\%\ for $\cte < 0.996$ as was shown in Ref. \cite{ich}. For two-photon 
final states $\cte$ is identical to $|\ct |$ and
for three-photon events in which the third photon is along the beam direction,
$\theta^{\ast}$ is equivalent to the scattering angle in the centre-of-mass 
system of the two observed photons. Since the angular definition is based on 
the two highest energy photons of the event, events with one of those escaping 
detection along the beam-axis are rejected from the analysis.

\section{Event selection}

Events are selected by requiring two or more clusters of energy in the 
electromagnetic calorimeter (ECAL). A cluster is selected
as a photon candidate if it is within the polar angle range 
$|\ct| < 0.97$. The cluster must consist of at least two lead-glass blocks 
and the cluster energy has to exceed 1 GeV (uncorrected for possible
energy loss in the material before the ECAL).

There are two major classes of background. 
The first class consists of events without primary charged 
tracks. Certain cosmic ray events, beam halo and the Standard Model 
process $\epem\to\bar{\nu}\nu\g\g$ contribute to this background.
The second class can be identified by the presence of primary charged tracks. 
Bhabha events, for example, have similar electromagnetic cluster 
characteristics to $\ggg$ events, but are distinguished by
the presence of tracks in the central tracking chambers.

\subsection*{Rejection of non-physics backgrounds}
A cosmic-ray particle can pass through the hadronic and electromagnetic
calorimeters without producing a reconstructed track in the
central tracking chambers. These particles do not cross the beam-axis.
Since the HCAL and ECAL have different radii, the resulting hits in the 
two detectors occur separated in azimuth. To remove this background, we reject 
events with  3 or more track segments found in the muon chambers.
In the case of fewer than three such track segments, the
event rejection depends on the highest energy HCAL cluster of the event.
Events are rejected if this HCAL cluster is separated from each of the photon
candidates by more than 10$^{\circ}$ in azimuth and has at least 1 GeV of
deposited energy in the case of one or two muon track segments or 15\%\ of
the observed ECAL energy if no muon track segments are found.

Another type of background consists of events with large occupancy in the ECAL
well localised in the detector. To reject these events, cuts on the extent 
of adjoining clusters are applied. An event is rejected if one of these
accumulations consists of more than 12 ECAL clusters or has an extent of
more than 0.4 rad in $\theta$ or 0.4 rad/$\sin{\theta}$ in $\phi$. 
After these cuts, the remaining non-physics background has low visible energy 
and becomes negligible after the kinematic selection described below.

\subsection*{Kinematic requirements}
The selection is based primarily on the requirement of small missing
energy and small missing transverse momentum. Selection variables based on the
measured angles and assuming three-body kinematics are used where possible.
In the case of four or more photons no constraints are available.

The event sample is divided into four classes \ca -- \dd\ which are 
distinguished by the number of photon candidates $N_{\gamma}$, the 
acollinearity angle $\acol$ and the aplanarity angle $\acop$:
\ba 
\acol & = & 180^{\circ} - \alpha_{12}  \\
\acop & = & 360^{\circ} - (\alpha_{12} + \alpha_{13} + \alpha_{23}),
\ea
where $\alpha_{ij}$ is the angle between clusters $i$ and $j$ and 
the two highest energy clusters are labelled 1, 2. 

All events having an acollinearity angle $\acol < 10^{\circ}$ (i.e., with the
two highest energy clusters almost collinear) are assigned to class \ca , 
independent of the number of photon candidates. For true $\eeggg$ events in 
this class, the sum of the two highest cluster energies $E_S^I = E_1 + E_2$ 
should be close to the centre-of-mass energy $\sqrt{s}$. Events in class \ca\ 
are selected if $E_S^I > 0.6 \sqrt{s}$.

Class \cb\ consists of acollinear events ($\acol > 10^{\circ}$) with exactly 
two observed photon candidates. Events of this class typically contain an
energetic photon that escapes detection near the beam-axis ($|\ct |>0.97$). 
If the polar angle of this photon is approximated as $|\ct| =1$, its energy, 
$E_{\rm lost}$, can be estimated from the angles of the observed clusters 
$\theta_1$ and $\theta_2$:
\be
E_{\rm lost} =  \sqrt{s} \; \left( 1 + \frac{\sin{\theta_1} + \sin{\theta_2}}
                       {|\sin{(\theta_1 + \theta_2)}|}\right)^{-1}  \; .
 \label{elost} \ee
The energy sum $E_S^{I\!I}$ is calculated by summing the two observed cluster 
energies and the lost energy:
\be  E_S^{I\!I}  =  E_1 + E_2 + E_{\rm lost} \; . \ee
The imbalance $\B$, defined as
\be
\B = (\sin{\theta_1}+\sin{\theta_2}) 
\left| \cos{\left(\frac{\phi_1-\phi_2}{2}\right)}\right| \; ,
\label{dphi}
\ee
provides an approximate measure of the scaled transverse momentum of the event.
Events in class \cb\ are selected if $\B < 0.2$, $E_S^{I\!I} > 0.6  \sqrt{s}$
and $E_{\rm lost}$ is less than both $E_1$ and $E_2$. This last requirement 
ensures that the two highest energy photons are those observed.

Classes \cc\ and \dd\ contain acollinear events ($\acol > 10^{\circ}$) having 
3 or more observed photon candidates. In this case the cluster energies
must be used in addition to the cluster angles in calculating the transverse 
and longitudinal momenta ($\pt$, $\pl$) of the system. Since a non-zero 
longitudinal momentum could result from an additional unobserved photon along 
the beam direction, the energy sum $E_S^{I\!I\!I}$ is calculated by adding 
$\pl$ to the cluster energies $E_i$: 
\be E_S^{I\!I\!I} =  \sum_{i=1}^{N_{\gamma}} E_i + \pl \; . \ee
Events in classes \cc\ and \dd\ are selected if $E_S^{I\!I\!I} > 0.6 \sqrt{s}$ 
and  $\pt < 0.1 \sqrt{s}$. Again, the lost energy along the beam-axis, 
now determined by $\pl$, must be smaller than the energies of the two highest 
energy clusters.

\begin{figure}[p]
   \begin{center} \mbox{
          \epsfxsize=13.0cm
           \epsffile{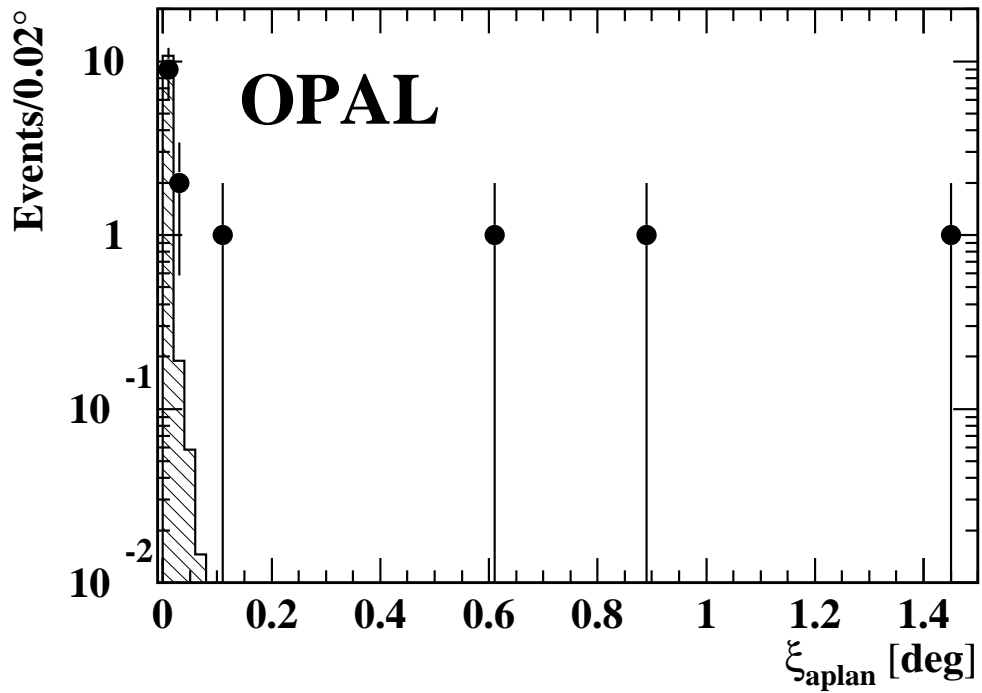}
           } \end{center}
\caption[ ]{The aplanarity for selected events with three or more photons
(classes \cc\ and \dd ). The points show the data and the histogram represents 
the $\O (\alpha^3)$ Monte Carlo expectation normalised to the integrated 
luminosity of the data. One additional event is observed at 
$\acop = 4.8^{\circ}$.}
\label{acopl}
\end{figure}

Figure \ref{acopl} shows the aplanarity distribution for selected events in
classes \cc\ and
\dd\ for data and for the $\O (\alpha^3)$ Monte Carlo after the charged-event 
rejection described in the next section. Most events are planar ($\acop < 0.1^{\circ}$)
and are well described by the Monte Carlo. They are consistent with exactly
three produced photons as simulated by the Monte Carlo. 
There are, however, 5 aplanar events ($\acop > 0.1^{\circ}$).
These events can be explained by the production of a fourth photon that 
escapes detection along the beam-axis in most cases. 
Planar events with exactly 3 observed photon candidates are assigned to
class \cc\ and aplanar three-photon events and events with more 
than three observed photon candidates are assigned to class \dd . 

The signal events are well separated from non-physics background 
in terms of energy and transverse momentum as has been demonstrated 
at lower energies \cite{ich}. The non-physics background is reduced to a 
negligible level after the above described kinematic requirements 
which are summarised in Table~\ref{cuts}.

\begin{table}[h]
\renewcommand{\arraystretch}{1.2}
\setlength{\tabcolsep}{1mm} 
\begin{center}
\begin{tabular}{|l|rcl|rcl|rcl|rcl|} \hline
Event class & \multicolumn{3}{c|}{\ca }&\multicolumn{3}{c|}{\cb }&
   \multicolumn{3}{c|}{\cc }&\multicolumn{3}{c|}{\dd}\\\hline
${\mbox{Number of}\atop\mbox{photon candidates}}$ \rule{0mm}{3.5ex}&
   $N_{\gamma}$ & $\geq$ & 2 & $N_{\gamma}$ & = & 2 & 
   $N_{\gamma}$ & = & 3 & $N_{\gamma}$   & $\geq$ & 3 \\
Acollinearity & 
   $\acol$& $<$ &$10^{\circ}$& $\acol$& $>$ &$10^{\circ}$ &
   $\acol$& $>$ &$10^{\circ}$& $\acol$& $>$ &$10^{\circ}$ \\
Aplanarity &
   & & & & & & $\acop$&$<$&$0.1^{\circ}$&$\acop $&$>$ &$0.1^{\circ}$ \\
Energy sum & 
   $E_S^I$& $>$ &$0.6 \sqrt{s}$ & $E_S^{I\!I}$& $>$ &$0.6 \sqrt{s}$ &
   $E_S^{I\!I\!I}$& $>$ &$0.6 \sqrt{s}$ & $E_S^{I\!I\!I}$& $>$ &$0.6 \sqrt{s}$ \\
Transverse momentum &
   & & & $\B$&$<$ &$0.2$ & 
   $\pt $& $<$ &$0.1  \sqrt{s}$ & $\pt $& $<$ &$0.1  \sqrt{s}$ \\
Longitudinal momentum &
   & & & $E_{\rm lost} $& $<$& $E_1 , E_2$ &
   $\pl $&       $<$ & $E_1 , E_2$ & $\pl $&       $<$ &$E_1 , E_2$ \\
\hline
\end{tabular}
\caption{Summary of the kinematic cuts. For definition of the variables see 
the text. In the case of more than three observed photons in class \dd\ 
there is no requirement on the aplanarity.}\label{cuts}\end{center}
\end{table}

\subsection*{Charged-event rejection}
The rejection of all events having tracks in the central tracking 
chambers would lead to an efficiency loss because of photon conversions. 
Nevertheless, contributions from any channel with primary charged tracks 
must be reduced to a negligible level. To achieve this,
events are rejected if they have a large number of hits in
the inner part of the drift chambers as described in Ref.~\cite{ich}.
In addition, events are rejected if there is a reconstructed track
separated by more than 10$^{\circ}$ in azimuth from all photon candidates. 
A good agreement of the efficiency and charged track rejection power has
been observed between data and Monte Carlo for different combinations 
of the vetoes. 

\section{Corrections and systematic errors}
\label{syserr}

Since the deviations from QED (Equations \ref{lambda} -- \ref{estar}) are given 
with respect to Born level, the observed angular distribution is corrected to 
Born level. The effect of radiative corrections is quantified by $\R$, the 
ratio of the angular distribution of the $\eeggg$ Monte Carlo and the 
Born-level cross-section:
\be \R = \xmc \!(\cte) \; \left/ \; \xb \right. . \label{corrad}\ee
The ratio $\R$ varies between 1.03 and 1.06 within the studied angular range 
and is used to correct the data bin by bin to Born level. A 1\%\ systematic 
error from higher-order effects is taken to be correlated between all bins.

The efficiency and angular resolution of the reconstruction are determined
using an $\O (\alpha^3)$ Monte Carlo sample with full detector simulation.
In the angular range $\cte < 0.87$ the efficiency varies smoothly between
90 -- 94\%\ and a polynomial parametrisation is used for the correction.
For the rest of the angular range the efficiency drops rapidly and
a bin-by-bin (bin width = 0.01) correction is made. The main reason for the 
loss is photon conversion. As a conservative estimate, the systematic error 
on the efficiency is taken to be a quarter of the inefficiency.
The resulting error ranges from 1.5\%\ ($\cte < 0.55$) to 15\%\ 
($\cte > 0.96$). The 1.5\%\ error is taken to be correlated between all bins.

The agreement between generated and reconstructed event angles is
very good; an event-angle resolution of $0.2^{\circ}$ is obtained.
In addition, the cluster angle has been compared to the
track angle for Bhabha events. For clusters with $|\ct | > 0.96$ the cluster 
angle is systematically about $0.4^{\circ}$ closer to the beam-axis than the 
track angle. For all other clusters the difference is less than
$0.1^{\circ}$. Due to the cut-off at $\cte <$ 0.97 this would lead
to a decrease of the measured total cross-section by 1.1\% 
if this effect is caused by the cluster angle. It is included in the
systematic error on the total cross-section.

The luminosity is derived from small-angle Bhabha scattering measured 
on both sides of the detector in the polar  angle region 
34 mrad $< \theta <$ 56 mrad. Uncertainties in the selection and in the 
theoretical cross-section, as well as a 30 MeV uncertainty on the beam energy 
lead to a systematic error of 0.5\%\ which is taken into account.

Table \ref{eventcuts} shows the number of selected events from the data and
from the expected Standard Model sources after the different
selection cuts. The preselected data have a large contribution from
non-physics backgrounds until the kinematic cuts are applied.
There is little  efficiency loss up to this point. The restriction on the 
missing longitudinal momentum ($E_{\rm lost}$ for class \cb ) rejects events 
with a high-energy photon escaping along the beam-axis.About 97\%\ of the 
remaining sample consists of Bhabha events and is well described by the 
Monte Carlo. After the charged-event rejection about one background event is 
expected which is negligible compared to the expected $\ggg$ signal.

\begin{table}[t]
\begin{center}
\renewcommand{\arraystretch}{1.3}
\renewcommand{\tabcolsep}{0.4em}
\begin{tabular}{|l|ccccccccc|} \hline
 Cut & Data & MC & $\ggg$ & $\rm e^+e^-(\gamma )$ & $\rm e\gamma (e)$ & $
\tau^+\tau^-$ &
 $\mu^+\mu^-$ & $\nu\overline{\nu}\g\g$ &$\rm q\overline{q}$ \\ \hline
 Preselection & 92833 & 37822 & 780 & 27778 & 3010 & 325 & 32 & 18 & 5879 \\
 Non-physics bg. reject. & 55791 & 32029 & 777 & 27657 & 2977 & 127 & 1.7 & 18 & 471 \\
 Kinematic req. & 28456 & 29263 & 769 & 26673 & 1760 &  32 & 0.6 & 1.1 & 27 \\ 
 $E_{\rm lost} , \pl < E_1 , E_2$ & 25284 & 26504 & 714 & 25707 & 62 & 8.7 & $<$0.1 &
                              $<$0.1 & 12 \\
 Charged event reject. &  620 & 603 & 602 & 0.3 & 0.6 & $<$0.1 & $<$0.1 &
                                             $<$0.1 & $<$0.1  \\ \hline
 \end{tabular}
\renewcommand{\arraystretch}{1.0}
 \caption{Number of selected events after the different cuts described in the 
text. The numbers are given for the data and the sum of Monte Carlo samples 
with the breakdown by final states given in the following columns. The BHWIDE 
generator is labelled by $\rm e^+e^-(\gamma )$ and TEEGG by $\rm e\gamma (e)$. 
All Monte Carlo predictions are normalised to the integrated luminosity of 
the data.}
 \label{eventcuts}
 \end{center}
 \end{table}

\section{Results}
\subsection*{Cross-section}

\begin{figure}[p]
   \begin{center} \mbox{
          \epsfxsize=13.0cm
           \epsffile{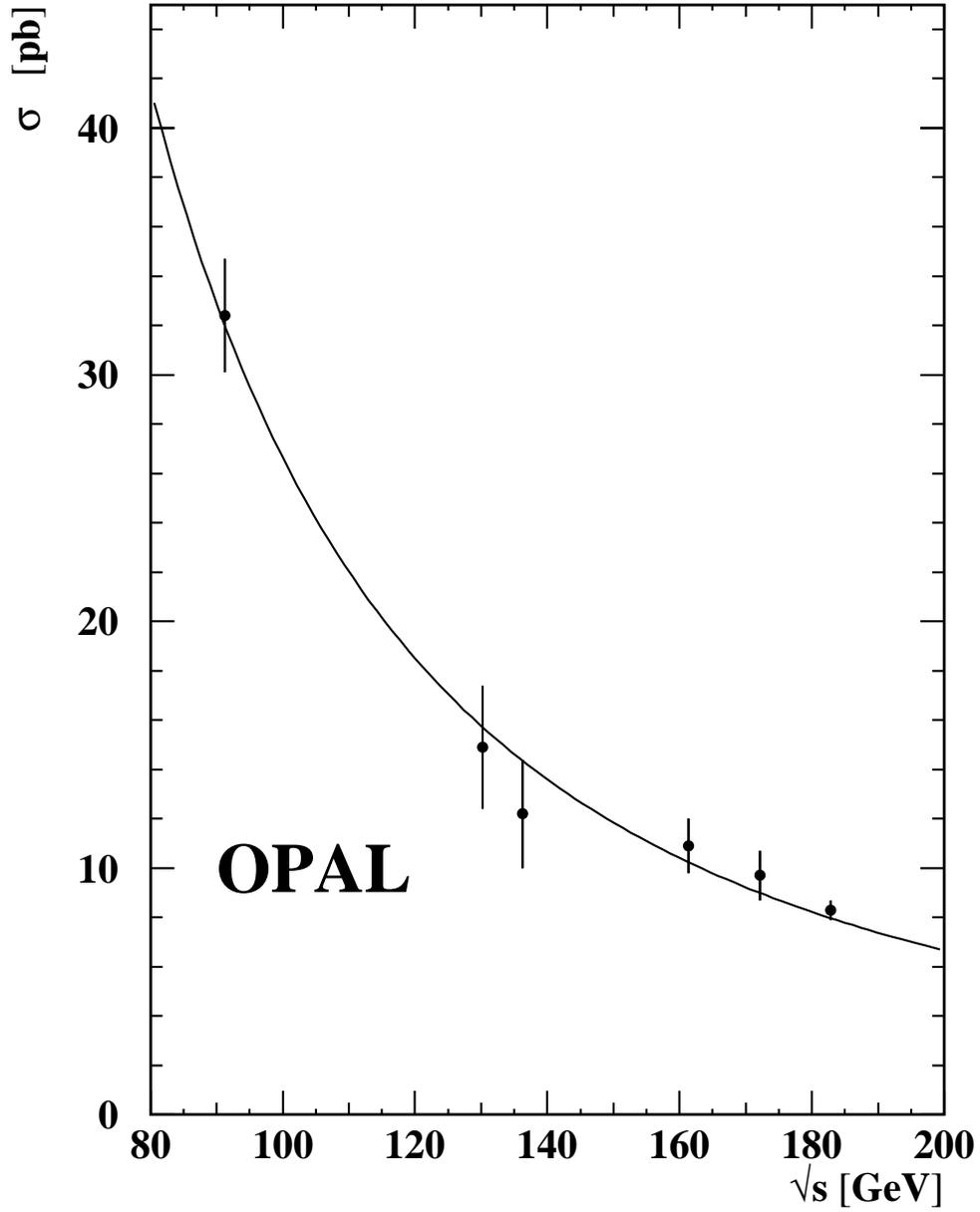}
           } \end{center}
\caption[ ]{Total cross-section for the process $\eegg$ with $\cte < 0.9$.
The data are corrected for efficiency loss and higher-order effects and
correspond to Born level. Results at lower energies are taken from
\cite{ich,opalt}. The curve shows the Born-level QED expectation.}
\label{totxsn}
\end{figure}

In Table \ref{events} the numbers of observed events in the different kinematic
classes are compared to the Monte Carlo expectation. Since there is no 
Monte Carlo generator for the case of a four-photon event where at least one 
photon is along the beam axis, no prediction for class \dd\ events with three 
observed photons is given. The prediction for events with
four or more observed photons is calculated using FGAM.
One event with four detected photons is observed.
The total cross-section $\sigma$ for the process $\eeggg$ determined
from 620 events selected in the range $\cte < 0.97$ is also given 
in Table \ref{events}. The total error is statistics dominated.
The cross-section is corrected for detection efficiency and $\O (\alpha^3)$
effects derived from $\R$ (Eq. \ref{corrad}).
The cross-section at the different LEP energies as measured by OPAL in 
the range $\cte < 0.9$ is shown in Figure \ref{totxsn}. All measurements 
are in good agreement with the QED prediction.

\begin{table}[b]
\begin{center}
\renewcommand{\arraystretch}{1.3}
\begin{tabular}{|l|c|c|c|c|c||c||c|c|}\hline
Class & \ca & \cb & \cc & \multicolumn{2}{c||}{ \dd } & all 
 & \multicolumn{2}{c|}{cross-section [pb]}\\
 $N_{\gamma}$ &$\geq 2$  & 2  & 3 & ~3~ & $\ge 4$ & events & $\cte < 0.9$ & $\cte < 0.97$\\
\hline
observed & 558 & 46 & 11 & 4  & 1   & 620 & $8.4 \pm 0.4$ & $12.9 \pm 0.7$\\
expected & 546 & 44.6 & 11.3 & -- & 0.42 & 602 & 8.0 & 12.5 \\ \hline
\end{tabular}
\caption[ ]{Number of observed and expected events in the angular range
$\cte <$ 0.97 for different classes. $N_{\gamma}$ is the number of observed photon
candidates. The total cross-section is corrected to Born level and is given for
two angular ranges to allow comparison of these results with those of
previous OPAL measurements. The errors are dominated by statistics.}
\label{events}
\end{center}
\end{table}

\begin{figure}[p]
   \begin{center}
      \mbox{
          \epsfxsize=14.0cm
          \epsffile{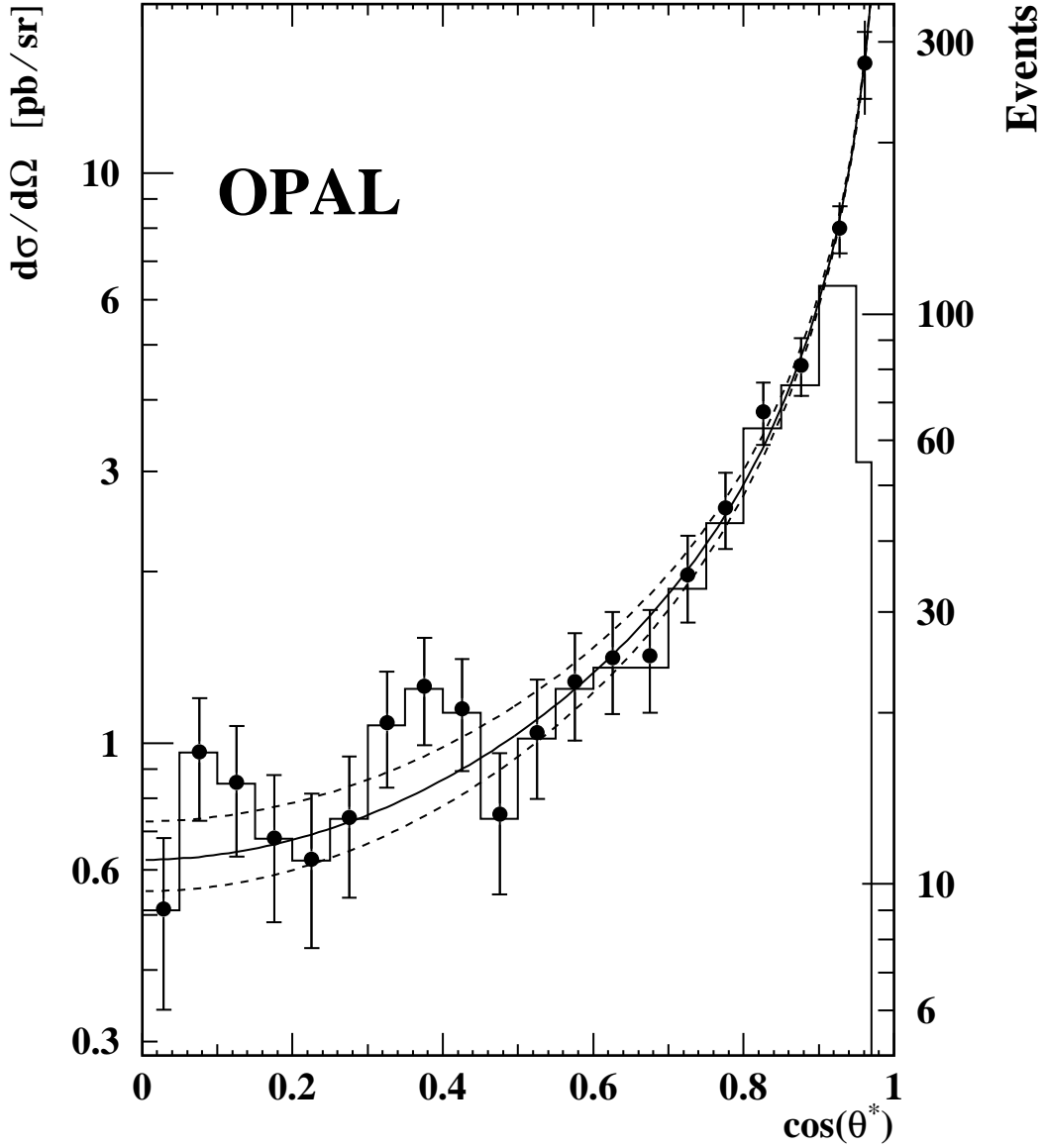}
           }
   \end{center}
\caption[ ]{The measured angular distribution for the process 
$\epem\to\g\g(\g)$ at $\sqrt{s} = $ 183 GeV. The histogram shows the observed 
number of events per bin. Note the smaller width of the highest $\cte$ bin.
The points show the number of events corrected for efficiency and radiative
effects. The inner error-bars correspond to the statistical error and the 
outer error-bars to the total error. The solid curve corresponds to
the Born-level QED prediction. The dashed lines represent the 95\% CL
interval of the fit to the function $\xl$.  }
\label{wq}
\end{figure}

The angular distribution of the observed events and the measured differential 
cross-section, obtained by applying efficiency and radiative corrections, are 
shown in Figure \ref{wq}. The data have a $\rm\chi^2 /NDF = 12.1/20$ with 
respect to the QED expectation (solid line). The 95\%\ CL interval of a 
$\chi^2$-fit to the data of the function $\xl$ (Eq. \ref{lambda}) is also 
shown (dashed line). To obtain the limits at 95\% confidence level the
probability is normalised to the physically allowed region,
i.e. $\Lambda_+ > 0$ and $\Lambda_- < 0$ as described in Ref. \cite{pdg}.
For both functions $\xl$ and $\xq$ the $\chi^2$ distribution is parabolic
as a function of the chosen fitting parameters $\Lambda_\pm^{-4}$ and 
$\Lambda'^{-6}$. The asymmetric limits 
$\lambda_{\pm}$ on the fitting parameter can be obtained by:
\be \frac{\int^{\lambda_+}_0 G(x,\mu ,\sigma ) dx}
         {\int^{\infty   }_0 G(x,\mu ,\sigma ) dx} = 0.95 \; 
    \hspace{7mm}\mbox{and}\hspace{7mm}
    \frac{\int_{\lambda_-}^0 G(x,\mu ,\sigma ) dx}
         {\int_{-\infty  }^0 G(x,\mu ,\sigma ) dx} = 0.95 \; , \ee
where $G$ is a Gaussian with the central value and error of the fit
result denoted by $\mu$ and $\sigma$, respectively. 

\begin{figure}[p]
   \begin{center}
      \mbox{
          \epsfxsize=13.0cm
          \epsffile{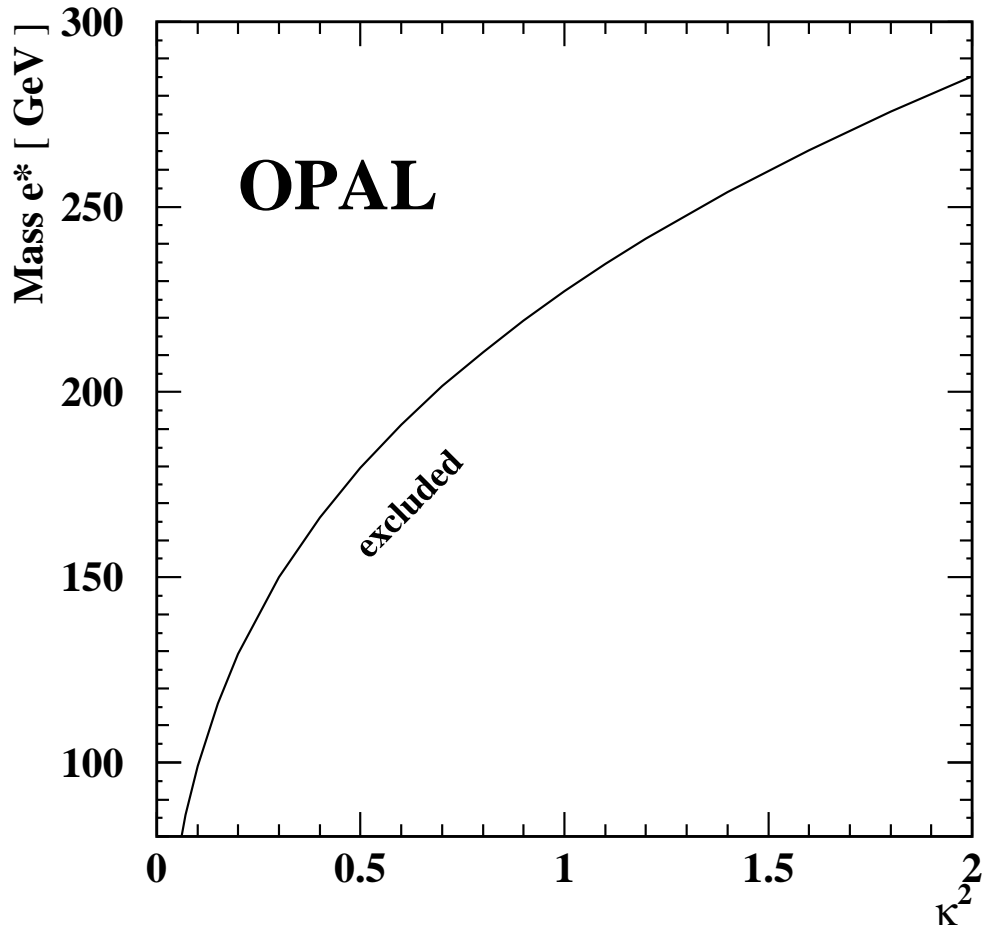}
           }
   \end{center}
\caption[ ]{Lower limit (95\%\ CL) on 
the mass $M_{\rm e^{\ast}}$ of an excited electron as a function of
the square of the $\rm e^{\ast}e\gamma$ coupling constant $\kappa^2$.}
\label{elimit}
\end{figure}

The limit on the mass of an excited electron $M_{\rm e^{\ast}}$ as a function
of the coupling constant $\kappa$ for the ($\mathrm{e^*e\g}$)-vertex, which 
is fixed during the fit, is shown in Figure \ref{elimit}. In the case of 
$\xe$ the cross-section does not depend linearly on the chosen fitting 
parameter $M_{\rm e^{\ast}}^{-2}$ and the limit corresponds to
an increase of the $\chi^2$ by 3.84 with respect to the minimum.

The fit results are summarised in Table \ref{result}. All results are from 
the data taken at 183 GeV. The inclusion of the data taken at 130 -- 172 GeV 
does not substantially improve the results.  The limits obtained are roughly 
40 GeV higher than our previous results. Similar limits are obtained by 
ALEPH and DELPHI at 183 GeV \cite{aleph183}.

\begin{table}[h]
\begin{center}
\renewcommand{\arraystretch}{1.5}
\begin{tabular}{|c|c|c|}\hline
Fit parameter & Fit result & 95\%\ CL Limit [GeV] \\ \hline
 &  & $\Lambda_+ >$  233   \\
 \raisebox{2.2ex}[-2.2ex]{$\Lpm^{-4}$} &
 \raisebox{2.2ex}[-2.2ex]{$(1.04 \pm 1.34)\cdot 10^{-10}$ GeV$^{-4}$}
 & $\Lambda_- > $ 265   \\
$\Lambda '^{-6}$ & $ (1.11 \pm 1.29 )\cdot
                                  10^{-17}$ GeV$^{-6}$
& $\Lambda '~ > $ 557   \\
$M_{\rm e^{\ast}}^{-2}$ & $ \left(1.06{+0.51 \atop -3.23}\right)\cdot
                                         10^{-5}$ GeV$^{-2} $
& $M_{\rm e^{\ast}} > $  227  \\ \hline
\end{tabular}
\renewcommand{\arraystretch}{1.0}
\caption[ ]{Fit results and 95\% CL lower limits obtained from the fit to the
differential cross-section.
The limit for the mass of an excited electron is determined
assuming the coupling constant $\kappa = 1$. }
\label{result}
\end{center}
\end{table}

\subsection*{Resonance production}

\begin{figure}[p]
   \begin{center} \mbox{
          \epsfxsize=14.0cm
           \epsffile{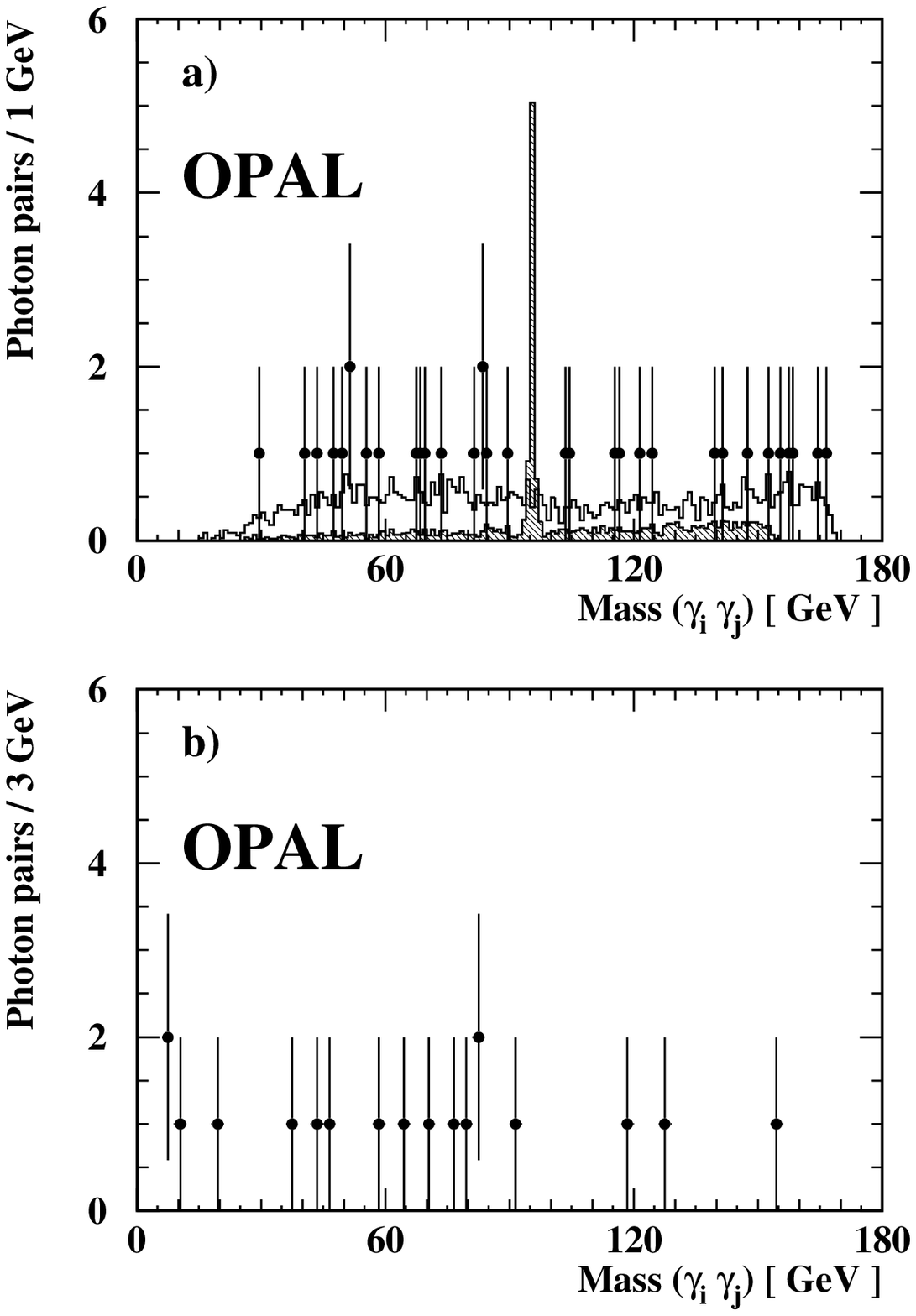}
           } \end{center}
\caption[ ]{The invariant mass of photon pairs from a) class-\cc\ and b)
class-\dd\ events. The points are the data and the open histogram the 
$\eeggg$ Monte Carlo expectation scaled by a factor of two for clarity.
The hatched histogram in a) represents a $\gamma\gamma$ resonance at 95.5~GeV
with a cross-section times branching ratio of 0.10 pb. In each case, the 
binning is chosen to match the expected mass resolution.}
\label{pmass}
\end{figure}

A resonance X produced by the process $\epem\to{\rm X}\g$ and decaying
into two photons, ${\rm X} \to \g\g$, could be seen in the two-photon invariant mass
spectrum, since this process leads to a three-photon final state without missing
energy. Searches for such a resonance have been performed previously at the 
Z$^0$ peak \cite{gres} and at higher energies \cite{ich,l3}, leading to bounds 
on Higgs and gauge boson interactions \cite{eboli2}.
For this search, 16 events from classes \cc\ and \dd\ are used.
The invariant mass of each photon pair is shown separately for class \cc\
(Figure \ref{pmass}a) and class \dd\ (Figure \ref{pmass}b).

For class \cc\ the energies of the three photons are not based on the measured 
cluster energies but are calculated from the photon angles assuming
three body kinematics:
\be
E_k \propto \sin{\alpha_{ij}} \; ; \;  E_1 + E_2 + E_3 = \sqrt{s} ,
\ee
with $E_k$ the energy of one photon and $\alpha_{ij}$ the angle between the
other two photons. In this case a typical mass resolution for photon pairs is
about 0.5 GeV. For class-\dd\ events the invariant mass is calculated from the 
measured cluster energies and angles with a typical mass resolution of 3 GeV.

\begin{figure}[p]
   \begin{center}
      \mbox{
          \epsfxsize=14.0cm
          \epsffile{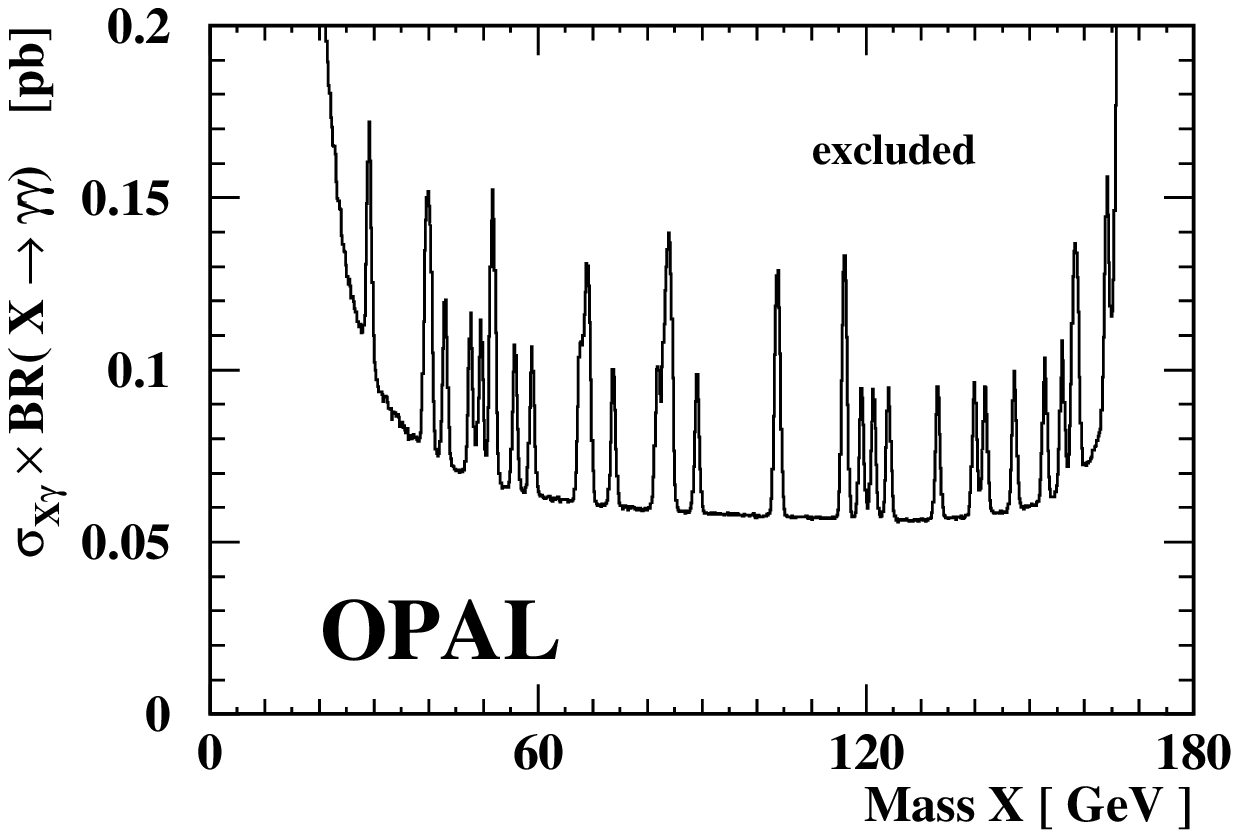}
           }
   \end{center}
\caption[ ]{Upper limit (95\%\ CL) for the cross-section times branching ratio 
for the process $\rm \epem\to X \g$, $\rm X\to\g\g$ as a function of the mass 
of the resonance X. Only class-\cc\ events are used for this result.}
\label{mlimit}
\end{figure}

The distribution obtained from class-\cc\ events is consistent with the 
Monte Carlo expectation from the QED process $\eeggg$ as seen in Figure 
\ref{pmass}a. In neither the class-\cc\ nor the class-\dd\ distribution is 
there evidence for an enhancement due to a resonance.
An upper limit on the total production cross-section multiplied by the
photonic branching ratio is calculated using the method of Ref. \cite{bock}. 
This method uses fractional event counting where the weights assigned to each
photon pair depend on the expected resolution and the difference between the 
hypothetical and the reconstructed mass. The limits shown in 
Figure \ref{mlimit} are obtained using only class-\cc\ events assuming the 
natural width of the resonance to be negligible. The $\eeggg$ background is 
subtracted. For the efficiency correction the production and subsequent decay 
of the resonance are assumed to be isotropic.
The mass range is limited by the acollinearity restriction.
Regarding a model with anomalous couplings of the Higgs boson \cite{eboli2}
this analysis gives access to a larger mass range than the analysis of 
$\epem\to\rm H Z$ with $\rm H \to \gamma\gamma$.

\section{Conclusions}

The process $\eeggg$ has been studied using data taken with the OPAL
detector at a centre-of-mass energy of 183 GeV.
The measured angular distribution and total cross-section for
this process both agree well with QED predictions. The
limits (95\%\ CL) on cut-off parameters are $\Lambda_+ > $ 233 GeV,
$\Lambda_- > $ 265 GeV and  $\Lambda ' > $ 557 GeV. An excited
electron is excluded for $M_{\rm e^{\ast}} >$ 227 GeV assuming the 
$\rm e^{\ast}e\gamma$ and $\rm ee\gamma$ coupling to be the same. In the 
$\gamma\gamma$ invariant mass spectrum using events with at least three 
final-state photons, no evidence is found for a resonance X decaying to $\g\g$.
A limit on the production cross-section times branching ratio is derived as a 
function of the mass $M_{\rm X}$. 
One event with four detected photons is observed.

\section{Acknowledgements}
We particularly wish to thank the SL Division for the efficient operation
of the LEP accelerator at all energies
 and for their continuing close cooperation with
our experimental group.  We thank our colleagues from CEA, DAPNIA/SPP,
CE-Saclay for their efforts over the years on the time-of-flight and trigger
systems which we continue to use.  In addition to the support staff at our own
institutions we are pleased to acknowledge the  \\
Department of Energy, USA, \\
National Science Foundation, USA, \\
Particle Physics and Astronomy Research Council, UK, \\
Natural Sciences and Engineering Research Council, Canada, \\
Israel Science Foundation, administered by the Israel
Academy of Science and Humanities, \\
Minerva Gesellschaft, \\
Benoziyo Center for High Energy Physics,\\
Japanese Ministry of Education, Science and Culture (the
Monbusho) and a grant under the Monbusho International
Science Research Program,\\
German Israeli Bi-national Science Foundation (GIF).

\clearpage

\begin{thebibliography}{20}
\bibitem{drell}S.D. Drell, Ann. Phys. {\bf 4} (1958) 75.
\bibitem{mcgen}F.A. Berends and R. Kleiss, Nucl. Phys. {\bf B186} (1981) 22.
\bibitem{eboli}O.J.P. \'{E}boli, A.A. Natale and S.F. Novaes, Phys. Lett. {\bf B271} (1991) 274.
\bibitem{litke}A. Litke, Ph.D.Thesis, Harvard University, unpublished (1970).
\bibitem{ich}OPAL Collaboration, K. Ackerstaff et al., Eur. Phys. J. {\bf C1} (1998) 21.
\bibitem{l3}L3 Collaboration, M. Acciarri et al., Phys. Lett. {\bf B413} (1997) 159.
\bibitem{aleph} ALEPH Collaboration, D. Buskulic et al., Phys. Lett. {\bf B384} (1996) 333.
\bibitem{delphi} DELPHI Collaboration, P. Abreu et al., Phys. Lett. {\bf B268} (1991) 296.
\bibitem{det}OPAL Collaboration, K. Ahmet et al., Nucl. Instr. and Meth. {\bf A305} (1991) 275.
\bibitem{fgam}CALCUL Collaboration, F.A. Berends et al., Nucl. Phys. {\bf B239} (1984) 395; \\
              P. Janot, PhD Thesis, LAL 87-31 (1987).
\bibitem{mcbh}S. Jadach et al., Phys. Lett. {\bf B390} (1997) 298.
\bibitem{mcte}D. Karlen, Nucl. Phys. {\bf B289} (1987) 23.
\bibitem{mctt}S. Jadach et al., Comp. Phys. Comm. {\bf 66} (1991) 276.
\bibitem{mcmh}T. Sj\"ostrand and M. Bengtsson, Comp. Phys. Comm. {\bf 43} (1987) 367; \\
              T. Sj\"ostrand, Comp. Phys. Comm. {\bf 39} (1986) 347.
\bibitem{mcdet}OPAL Collaboration, J. Allison et al., Nucl. Instr. and Meth. {\bf A317} (1992) 47.
\bibitem{bg}I. Harris and L.M. Brown, Phys. Rev. {\bf 105} (1957) 1656; \\
            F.A. Berends and R. Gastmans, Nucl. Phys. {\bf B61} (1973) 414.
\bibitem{pdg}Review of Particle Physics, R.M. Barnett et al., Phys. Rev.  {\bf D54} (1996) 164.
\bibitem{aleph183} ALEPH Collaboration, R. Barate et al., CERN-EP/98-053, submitted to Phys. Lett. B; \\
                  DELPHI Collaboration, P. Abreu et al., CERN-EP/98-075, submitted to Phys. Lett. B. 
\bibitem{gres} OPAL Collaboration, P.D. Acton et al., Phys. Lett. {\bf B311} (1993) 391; \\
               L3 Collaboration, M. Acciarri et al., Phys. Lett. {\bf B345} (1995) 609.
\bibitem{eboli2}O.J.P. \'{E}boli, M.C. Gonzalez-Garcia, S.M. Lietti and S.F. Novaes, hep-ph/9802408 (1998).
\bibitem{bock}P. Bock, Heidelberg preprint HD-PY 96/05,
     submitted to Nucl. Instrum. Meth. .
\bibitem{opalt} OPAL Collaboration, M.Z. Akrawy et al., Phys. Lett {\bf B257} (1991) 531.
\end{thebibliography}
\end{document}